\documentclass[12pt]{article}
\usepackage[a4paper,left=1.5748in,right=0.59055in,top=1.5748in]{geometry}
\usepackage{amssymb}
\usepackage{setspace}
\usepackage{amsmath}
\usepackage{enumerate}

\newtheorem{thm}{Theorem}[section]
\newtheorem{prop}[thm]{\textsc{Proposition}}
\newtheorem{lem}[thm]{\textsc{Lemma}}
\newtheorem{cor}[thm]{\textsc{Corollary}}
\newtheorem{exam}[thm]{\textsc{Example}}

\newtheorem{rem}[thm]{\textsc{Remark}}

\newcommand{\pf}{\noindent {\bf Proof.  }}
\newcommand{\qed}{\hfill $\Box$ \\}

\def\0{{\mathbf 0}}
\def\C{{\mathcal C}}

\newcommand{\F}{\mathbb{F}}

\begin{document}

\title{ On linear complementary-dual multinegacirculant codes}
\author{Adel Alahmadi\thanks{Math. Dept., King Abdulaziz University, Jeddah, Saudi Arabia, {Email: \tt adelnife2@yahoo.com}}, Cem G\"{u}neri\thanks{Sabanc\i \ University, FENS, 34956 Istanbul, Turkey, {Email: \tt guneri@sabanciuniv.edu}}, Buket \"{O}zkaya\thanks{ Sabanc\i \ University, FENS, 34956 Istanbul, Turkey,
{Email: \tt buketozkaya@sabanciuniv.edu}},\\ Hatoon Shoaib\thanks{Math.
Dept., King Abdulaziz University, Jeddah, Saudi Arabia, {Email: \tt
hashoaib@kau.edu.sa }}, Patrick Sol\'e\thanks{ CNRS/LAGA,
Universit\'e de Paris 8, 93 526 Saint-Denis, France, {Email: \tt
sole@math.univ-paris13.fr}} }
\date{}
\maketitle
\begin{abstract} Linear codes with complementary-duals (LCD) are linear codes that intersect with their dual trivially.
Multinegacirculant codes of index $2$ that are LCD are characterized algebraically and some good codes are found in this family. Exact
enumeration is performed for indices 2 and 3, and for all indices $t$ for a special case of the co-index by using their concatenated structure.
Asymptotic existence results are derived for the special class of such codes that are one-generator and have co-index a power of two by means of
Dickson polynomials. This shows that there are infinite families of LCD multinegacirculant codes with relative distance satisfying a modified
Varshamov-Gilbert bound.
\end{abstract}

\vspace*{1cm}


\section{Introduction}
Linear complementary dual codes (LCD) are linear codes that intersect with their dual trivially. They were introduced by Massey in \cite{Ma},
and rediscovered a few years ago in the context of side-channel attacks \cite{CG}, and recently, in the domain of quantum error correcting codes
for entanglement assisted communication \cite{GJG}. In a recent paper the class of quasi-cyclic LCD codes was shown to be ``good" \cite{GOS}.
The main result of the present paper is to show  that some classes of one-generator quasi-twisted codes, an odd characteristic analogue of
quasi-cyclic codes, are not only good, but better than the Varshamov-Gilbert bound. Some exact enumeration results for index $2$ and index $3$
are also derived. For a general index $t$ and co-index power of 2, a special enumeration is given which is needed for asymptotic analysis. A
construction technique for index $2$ (double negacirculant codes), and some examples of optimal or best such codes in modest lengths are given.
The main technical ingredients of the proofs are some results on the number of solutions of certain diagonal equations over finite fields, given
in the Appendix.

The material is organized as follows. The next section surveys the algebraic structures of the codes we study. Section 3 collects the notions
and notations needed in the rest of the paper. Section 4 recalls some known facts on double negacirculant codes.  Section 5 describes the
factorization  of $x^n+1$ over $\F_q$ when $n$ is a power of $2,$ and $q$ is odd. Section 6 characterizes algebraically LCD double negacirculant
codes and gives some examples with optimum distance in modest lengths. Section 7 contains exact enumeration formulae. Section 8 builds on
Section 7 to study the asymptotic performance of multinegacirculant codes of index $t$ where $ t \geq 2$.
 Section 9 recapitulates the results we obtained, and exhibits some challenging open problems.
Section 10 is the Appendix mentioned above.

\section{Preliminaries}
A matrix $A$ over a finite field $\F_q$ is said to be {\em
negacirculant} if its rows are obtained by successive negashifts
from the first row. A {\em negashift} maps the vector
$(x_0,\dots,x_{n-1})\in \F_q^n$ to $(-x_{n-1},x_0,\dots,x_{n-2}).$

In this paper we consider {\em double negacirculant} (DN) codes over finite fields, that is, $[2n,n]$ codes with generator matrices of shape
$(I,A)$ with $I$ the identity matrix of size $n$ and $A$ a negacirculant matrix of order $n.$ This construction was introduced in \cite{HG}
under the name {\em quasi-twisted code}. We prefer to reserve this term for the more general class of codes described in \cite{J}. Before
describing this class, we also define negacirculant codes of higher index over finite fields.

Three-negacirculant codes are $[3n,n]$ codes with generator matrices of the shape $(I,A,B)$ with $A,B$ negacirculant matrices of order $n$.
Similarly one can define the $t$-negacirculant code (for any $t>3$) and call all such codes multinegacirculant.

A code of length $N$ is {\em quasi-twisted} of index $\ell$ where
$\ell \mid N,$ and co-index $m=\frac{N}{\ell}$ if it is invariant
under the power $T_{\alpha}^\ell$ of the {\em constashift}
$T_{\alpha}$ defined as

$$T_{\alpha}: (x_0,\dots,x_{N-1}) \mapsto (\alpha x_{N-1},x_0,\dots,x_{N-2}).$$
Thus for $\alpha=-1,$ DN codes are quasi-twisted codes of index $2,$ and three-negacirculant codes are quasi-twisted codes of index $3$, etc.
Such a code affords a natural module structure over the auxiliary ring
$$R(m,\F_q)=\frac{\F_q[x]}{\langle x^m-{\alpha}\rangle }.$$ In other words, it can
be regarded as a code of length $\ell$ over the ring $R(m,\F_q).$ When this module has one generator over that ring the code is said to be {\em
one-generator}. An algebraic way to study such a code is to decompose the semilocal ring $R(m,\F_q)$ as a direct sum of local rings by the
Chinese Remainder Theorem \cite{J}, thus following the approach initiated for quasi-cyclic codes in \cite{LS}. The benefit of this technique is
to reduce the study of QT codes to that of shorter codes over larger alphabets. Besides, the study of duality is made transparent, thus allowing
the construction of LCD QT codes, as in, for instance, \cite{J}. The number of rings occurring in the decomposition of $R(m,\F_q)$ equals the
number of irreducible factors of $x^m-\alpha.$ In the following section we focus on LCD DN codes. These have been explored numerically in
\cite{GOS}. For some specific alphabets, it can be shown that in that case $x^n+1$ can be factored into the product of two irreducible
polynomials \cite{LN,M}. This is a favourable situation in which to apply the Chinese Remainder Theorem approach of \cite{LS,J}, as the
decomposition of $R(m,\F_q)$ contains only two terms. It allows to derive exact enumeration formulae and, from there, using the so-called
expurgated random coding technique, to give an asymptotic lower bound on the minimum distance of these one-generator quasi-twisted codes of
fixed arbitrary index. This is an analogue of the Varshamov Gilbert bound.

\section{Definitions and Notation}
\subsection{Codes}
Let $\F_q$ denote the finite field of order some prime-power $q.$ We assume throughout that $q$ is odd. In the following, we shall consider
codes over $\F_q$ of length $2n$ which is coprime to $q$. Their generator matrices $G$ will be of the form $G=(I,A),$ where $I$ is the identity
matrix of order $n$ and $A$ is an $(n\times n)$-negacirculant matrix. We call such codes {\em double negacirculant} (DN) codes. We shall denote
by $\C_a$ the DN code with first row of $A$ being the $x-$expansion of $a(x)$ in the ring $R(n,\F_q).$ Specifically, if $a(x)=a_0+a_1x+ \cdots
+a_{n-1}x^{n-1}$, then the first row is $(a_0,a_1,\ldots , a_{n-1})$ and the other rows of $A$ are obtained by negashifts of the previous row.

If $\C(m)$ is a family of codes of parameters $[m,k_m,d_m],$ the
{\em rate} $R$ and {\em relative distance} $\delta$ are defined as
$$R=\limsup_{m \rightarrow \infty}\frac{k_m}{m},$$
and
$$\delta=\liminf_{m \rightarrow \infty}\frac{d_m}{m}.$$
Both limits are finite as they are limits of bounded quantities. Such a
family of codes is said to be {\it good } if $R\delta \neq 0.$

Recall the $q-$ary {\em entropy function} is defined for $0<y<
\frac{q-1}{q}$ by $$
H_q(y)=y\log_q(q-1)-y\log_q(y)-(1-y)\log_q(1-y).$$ This quantity is
instrumental in the estimation of the volume of high-dimensional
Hamming balls when the base field is $\F_q.$ The result we are using
is that the volume of the Hamming ball of radius $yn$ is, up to
subexponential terms,  $q^{nH_q(y)},$ when $0<y<1,$ and $n$ goes to
infinity \cite[Lemma 2.10.3]{HP}.

\subsection{Polynomials}
The {\em Dickson polynomials} (of the first kind)  are given by
$D_0(x,\alpha) = 2$, and for $m > 0,$ by

  $$  D_m(x,\alpha)=\sum_{p=0}^{\lfloor m/2\rfloor}\frac{m}{m-p} \binom{m-p}{p} (-\alpha)^p x^{m-2p}. $$
  The $D_m$ satisfy the identity

   $$ D_m(u + \alpha/u,\alpha) = u^m + (\alpha/u)^m .$$

\section{Background on Double Negacirculant Codes}
We consider double negacirculant (DN) codes over finite fields. These are $[2n,n]$ codes over $\F_q$, where the codewords are closed under two
negashifts. DN codes have systematic generating matrices $G=(I_n:A)$ with $A$ an $n\times n$ negacirculant matrix. Algebraically, we can view
such a code as an $R$-module in $R^2$, generated by $(1,a(x))$, where $R= \F_q[x]/\langle x^n+1 \rangle$. In other words, a DN code is an index
2 quasi-twisted code with $\lambda=-1$ (see \cite{J} for notation and more information on quasi-twisted codes).

If the characteristic of $\F_q$ is 2, then a cyclic shift and a
negashift are the same. Hence a DN code is simply a double circulant
code. Therefore we assume throughout that $q$ is odd. Moreover, we
will assume that $n$ is relatively prime to $q$.

As in \cite{J}, assume that the factorization of $x^n+1$ into
irreducible polynomials over $\F_q$ is of the form
\begin{equation}\label{factorization}
x^n+1=\alpha \prod_{i=1}^s g_i(x) \prod_{j=1}^t h_j(x)h^*_j(x),
\end{equation}
where $\alpha \in \F_q$, $g_j$ a self-reciprocal polynomial and $*$
denotes reciprocation. Let $\xi$ be a primitive $(2n)^{th}$ root of
unity over $\F_q$. Then $\xi^n=-1$ and hence $\xi$ is a root of
$x^n+1$. Moreover, $\xi^{-1}=-\xi^{n-1}$. Assume that
$g_i(\xi^{u_i})=0$ and $h_j(\xi^{v_j})=0$ (for all $i,j$). Then we
also have $h^*_j(-\xi^{(n-1)v_j})=0$. By the Chinese Remainder
Theorem (CRT) we have
\begin{eqnarray*}
R & \simeq & (\bigoplus_{i=1}^s \F_q[x]/\langle g_i \rangle )\oplus
(\bigoplus_{j=1}^t (\F_q[x]/ \langle h_j \rangle \oplus (\F_q[x]/\langle h^*_j \rangle)) \\
 &=&(\bigoplus_{i=1}^s \F_q(\xi^{u_i}))\oplus (\bigoplus_{j=1}^t (\F_q(\xi^{v_j}) \oplus (\F_q(-\xi^{(n-1)v_j}))).
 \end{eqnarray*}
Let $G_i= \F_q[x]/\langle g_i \rangle$, $H'_j=\F_q[x]/\langle h_j
\rangle$ and $H''_j=\F_q[x]/\langle h^*_j \rangle$ for simplicity.
Note that all of these fields are extensions of $\F_q$. This
decomposition naturally extends to $R^2$ and then a linear code $\C \subset R^2$
decomposes as
\begin{equation}\label{fact}
\C= (\bigoplus_{i=1} ^s \C_i) \oplus (\bigoplus_{j=1} ^t (\C'_j
\oplus \C''_j)),
\end{equation}
where each component code (constituent) is a length $2$ linear code
defined over the respective base field $G_i,H'_j,H''_j$. More
specifically, again by CRT, we have
\begin{eqnarray}\label{consts}
\C_i&=&
\mbox{Span}_{G_i}\{ (1,a(\xi^{u_i}))\}, \ 1\leq i \leq s, \nonumber\\
\C'_j&=& \mbox{Span}_{H'_j}\{ (1,a(\xi^{v_j}))\}, \ 1\leq j \leq t,\\
\C''_j&=& \mbox{Span}_{H''_j}\{ (1,a(-\xi^{(n-1)v_j}))\}, \ 1\leq j \leq t. \nonumber \end{eqnarray} The Euclidean dual of $C$ in $\F_q^{2n}$ is
also a DN code and its decomposition is as follows (\cite[Theorem 3]{J}):
\begin{equation}\label{fact2}
\C^\perp= (\bigoplus_{i=1} ^s \C_i^{\perp_{H}}) \oplus
(\bigoplus_{j=1}^t (\C_j^{''\perp_E} \oplus \C_j^{'\perp_E})).
\end{equation}
Here, $\perp_{H}$ denotes the Hermitian dual on $G_i$ for all $1 \leq i \leq s$, and $\perp_E$ denotes the Euclidean dual on $H_j^{'}$ ,
$H_j^{''}$ for all $1 \leq j \leq t$. For instance,
$$(1,a(\xi^{u_i}))\cdot_{G_i}(1,a(\xi^{u_i}))=1+a(\xi^{u_i})a(-\xi^{(n-1)u_i}).$$

\begin{rem}
{\em Let us note that the CRT decomposition described in this section extends naturally to higher indices $t\geq 3$. In this case, the codes are
$R$-submodules of $R^t$.}
\end{rem}

\section{Factorizations}
The complete factorization of $x^{2^n}+1$ over $\F_q$ with $q \equiv
3$ (mod 4) is given by the following theorem \cite{M}.
\begin{thm}\label{1}
Let $q \equiv 3 \pmod{4},$ where $q=2^Am-1$, $A \geq 2$, $m$ an odd integer. Let $n \geq 2.$\\
(a) If $n < A$, then $x^{2^n}+1$ is the product of $2^{n-1}$
irreducible quadratic trinomials over $\F_q$
$$x^{2^n}+1= \prod_{\gamma \in \Gamma}(x^2+ \gamma x +1),$$
where $\Gamma$ is the set of all roots of $D_{2^{n-1}}(x,1)$.\\
(b) If $n \geq A$, then $x^{2^n}+1$ is the product of $2^{A-1}$
irreducible trinomials over $\F_q$
$$ x^{2^n}+1 = \prod_{\delta \in \Delta}(x^{n-A+1}+ \delta x^{n-A}-1),$$
where $\Delta$ is the set of all roots of $D_{2^{A-1}}(x,-1)$ in
$\F_q.$
\end{thm}

\begin{exam}
{\rm If $q=3$ i.e. $q \equiv 3 \pmod{4},$ then $q=2^{2}.1-1$ implies that $A=2$, $m=1$, and $\Delta=\{1,2\}$, then by Theorem \ref{1}:
$$x^{2^{n}}+1=(x^{2^{n-1}}+x^{2^{n-2}}+2)(x^{2^{n-1}}+2x^{2^{n-2}}+2).$$}
\end{exam}

\begin{thm}\label{2}
 Let $q \equiv 1 \pmod{4},$ where $q=2^Am+1$, $A \geq 2$, $m$ is odd integer. Denote by $U_k$ the set of all primitive $2^k$th roots of unity in $\F_q.$ If $n \geq 2$, then\\
 \begin{itemize}
  \item[(a)] If $n \leq A$, then $ord_{2^{n+1}}(q)=1$ and $x^{2^n}+1$ is the product of $2^n$ linear factors over $\F_q$
$$x^{2^n}+1 = \prod_{u \in U_{n+1}}(x+u).$$
  \item[(b)] If $n \geq A+1$, then $ord_{2^{n+1}}(q)=2^{n-A}$ and $x^{2^n}+1$ is the product of $2^A$ irreducible binomials over $\F_q$ of degree $2^{n-A}$
$$ x^{2^n}+1= \prod_{u \in U_{A+1}}(x^{2^{n-A}}+ u).$$

 \end{itemize}
\end{thm}

\begin{exam}
{\rm If $q=5$ i.e. $q \equiv 1$ (mod 4), then $q=2^{2}.1+1$ implies that $A=1$, $m=1$, and $U_2=\{2,3\}$ , then by Theorem \ref{2}:
$$x^{2^{n}}+1=(x^{2^{n-1}}+2)(x^{2^{n-1}}+3).$$}
\end{exam}

\section{Constructions and Examples}

In this section, we characterize linear complementary-dual ($\C\cap
\C^\perp=\{0\}$) DN codes. The proof is very similar to the double
circulant case (\cite[Theorem 5.1]{GOS}).

\begin{thm}\label{cover}
 Let $\C=\langle (1,a(x))\rangle \subset R^2$ be a double negacirculant code over $\F_q$. Then,\\
 $\C$ is linear complementary-dual if and only if $\gcd \left(1+a(x)a(-x^{n-1}),x^n+1\right)=1$.
\end{thm}

\pf Let $\xi$ be a primitive $(2n)^{th}$ root of unity, and assume that $x^n+1$ factors as in
(\ref{factorization}). Constituents of $\C$ are described in
(\ref{consts}). Note that each constituent is a 1-dimensional space
in the two dimensional ambient space. Hence, the dual
of any constituent is 1-dimensional too.

Note that in terms of constituents, we have that $\C$ is linear
complementary-dual if and only if $\C_i$ is linear
complementary-dual relative to Hermitian product in $G_i^2$ for all
$i$, and $\C'_j \cap \C_j^{''\perp}=\{0\}$, $\C''_j \cap
\C_j^{'\perp}=\{0\}$ for all $j$.

Observe that $\C_i \cap \C_i^{\perp_{G_i}} \neq \{0\}$ if and only
if $\C_i= \C_i^{\perp_{G_i}}$, which is equivalent to
$$ 1+a(\xi^{u_i})a(-\xi^{(n-1)u_i})=0.$$
On the other hand, $\C'_j \cap \C_j^{''\perp} \neq \{0\}$ if and
only if $\C'_j= \C_j^{''\perp}$, which is equivalent to
$$ 1+a(\xi^{v_j})a(-\xi^{(n-1)v_j})=0.$$
The last intersection $\C''_j \cap \C_j^{'\perp} \neq \{0\}$ does not bring a new condition. Therefore, being LCD for $\C$ is equivalent to the
polynomial $a(x)a(-x^{n-1})+1$ not vanishing at any root of $x^n+1$. \qed

The following table displays the best possible distances for double
negacirculant LCD codes $\C=\langle (1,a(x))\rangle \subset R^2,$
where $\C$ is DN of length $2n$ and dimension $n$ and index $2$. The
search was done in Magma (\cite{BCP}) for $q=5$ and random
$a(x) \in R$ satisfying the conditions in the above Theorem. Entries marked with $*$ are optimal or have best-known parameters.\\
\begin{center}
\begin{tabular}{|c||c|c|c|c|c|c|c|c|c|c|c|c|}
  \hline
  \textit{\textbf{n}} & 3 &4& 6&  7 &8 & 9 & 11 &12 & 13 & 14 & 16 & 17   \\
  \hline
  \textit{\textbf{d}} & 4* & 4* &6* &6* & 7* & 7* & 8* & 9* & 9 & 9 & 11* & 11* \\
  \hline
  \textit{\textbf{d*}} & 4 & 4 & 6 & 6 & 7 & 7 & 8 & 9 & 10 & 11 & 11  & 11  \\
  \hline
  \textit{\textbf{r}} & 480 & 128 & 384 & 28 & 64 & 36 & 44 & 48 & 52 & 56 & 64  & 68  \\
  \hline
\end{tabular}
\end{center}
Here, $\emph{d}$ is the minimum distance, $\emph{d*}$ is the highest minimum distance of a linear code of given length and dimension \cite{G},
and $\emph{r}$ is the size of automorphism group.

\section{Enumeration}
In this section we use repeatedly the following observation. If a quasi-twisted code has one generator as a module over the ring $R(m,\F_q),$
then it is either self-orthogonal or LCD.

\subsection{Index 2}
We give a general enumeration formula that is not needed for asymptotics, but of interest in its own right.
Recall that the so-called \emph{quadratic character} $\eta$ of $\F_q$ is defined as $\eta(x)=1$ if $x \in \F_q$ is a nonzero square and $\eta(x)=-1$ if not.
\begin{prop}\label{enum1}
 Let $q$ be an odd prime power, and $n \ge 1$ be an integer coprime to $q$. Assume that the factorization of $x^n+1$ into irreducible polynomials over $\F_q$ is of the form
 $$x^n+1=\alpha \prod_{i=1}^s g_i(x) \prod_{j=1}^t h_j(x)h^*_j(x),$$
with $\alpha \in \F_q^*$, and $g_i$ a self-reciprocal polynomial of degree $2d_i$, the polynomial $h_j$ is of degree $e_j$ and $*$ denotes
reciprocation. If $n$ is odd, then let $g_1=x+1$. The number of LCD double negacirculant codes over $\F_q$ of length $2n$ is
\begin{itemize}
  \item[$\bullet$] $(q-2)\prod_{i=2}^s (q^{2d_i}-q^{d_i}-2) \prod_{j=1}^t (q^{e_j}-1)(q^{e_j}-2)$ if $ \eta(-1)=1$ and,
  \item[$\bullet$] $q\prod_{i=2}^s (q^{2d_i}-q^{d_i}-2) \prod_{j=1}^t (q^{e_j}-1)(q^{e_j}-2)$ if $ \eta(-1)=-1$,
 \end{itemize}
if $n$ is odd and it is
$$\prod_{i=1}^s (q^{2d_i}-q^{d_i}-2) \prod_{j=1}^t (q^{e_j}-1)(q^{e_j}-2),$$
when $n$ is even.
\end{prop}

\pf We use the Chinese Remainder Theorem (CRT) decomposition of
$R(n,\F_q),$ as explained in $\S4$. Since we are counting
LCD quasi-twisted codes of index 2, we are reduced to count certain
codes of length 2 and dimension 1 over some extension $\F_Q$ of
$\F_q.$

In the case of $Q=q$ which happens for $n$ odd we have to count Euclidean self-orthogonal codes of length $2$ and dimension one over $\F_q.$ They are of the form
$\langle[1,a]\rangle,$ with $a$ a square root of $-1.$ We thus have $q$ or $q-2$ coefficients $a$ giving LCD codes, depending on $\eta(-1)=-1$ or $\eta(-1)=1.$

A {\it self-reciprocal} factor $g_i(x)$ of degree $2d_i$ leads to counting LCD hermitian codes of length $2$ over $\F_Q,$ where $Q=q^{2d_i}.$
The Hermitian self-dual codes of length 2 and dimension 1 over  $\F_Q$ are of the form $\langle[1,a]\rangle,$ with $a \in \F_Q,$ a solution of
$1+a^{1+\sqrt{Q}}=0.$ By finite field theory, this equation in $a$ admits $q^{d_i}+1$ roots in $\F_{q^{2d_i}}.$ The number of LCD codes sought
for is then $(q^{2d_i}-1)-(q^{d_i}+1).$ Note that the number of linear codes of length 2 over some $\F_Q$ admitting, along with their dual, a
systematic form is $Q-1$, all of dimension 1. We are thus excluding the code $\langle[1,0]\rangle,$ of dual $\langle[0,1]\rangle.$

In case of {\it reciprocal pairs} $(h'(x),h''(x))$, we need two codes of length 2 over $\F_Q$ that is, $C'=\langle[1,a']\rangle$, and $C''=
\langle[1,a'']\rangle^\perp, $ satisfying the condition that both $C' \cap C''^\bot$ and $C'' \cap C'^\bot$ are trivial . This boils down to the
condition $a' a'' \neq -1.$ So we have $q^{e_j}-1$ choices for $a',$ and $ a''\in H''_j \setminus \{0,- \frac{1}{a'}\}.$ This gives $q^{e_j}-2$
choices for $a''.$ Hence, in total we obtain $(q^{e_j}-1)(q^{e_j}-2)$ choices. \qed

We assume now that $q$ is such that $x^n+1,$ for $n$ a power of $2,$ has only two irreducible factors over $\F_q,$ say $h'(x)$ and $h''(x)$, and
that they are reciprocals of each other. Thus, $x^n+1=h'(x)h''(x).$ For convenience, let $K'=\frac{\F_q[x]}{\langle h'(x)\rangle}$ and
$K''=\frac{\F_q[x]}{\langle h''(x)\rangle}.$ These two fields are both isomorphic to $\F_{q^{n/2}}.$ By Theorems \ref{1} and \ref{2}, this is
the case if $q=4m\pm 1,$ with $m$ odd. For instance this happens if $q=3,5$ but not if $q=7.$ The following will be used in our asymptotic study
and it is a consequence of Proposition \ref{enum1}.

\begin{cor}\label{enuma}
 Let $q$ be odd, and $n$ be a power of $2$. If  $x^n+1$ factors as a product of two irreducible polynomials over $\F_q$, then the number of LCD double negacirculant codes over $\F_q$ of length $2n$ is $(q^{\frac{n}{2}}-1)(q^{\frac{n}{2}}-2)$.
\end{cor}

\subsection{Index 3}

\begin{prop}\label{enum2}
 Let $q$ be odd, and $n$ be coprime to $q$. Assume that the factorization of $x^n+1$ into irreducible polynomials over $\F_q$is of the form
 $$x^n+1=\alpha \prod_{i=1}^s g_i(x) \prod_{j=1}^t h_j(x)h^*_j(x),$$
with $\alpha \in \F_q^*$, and $g_i$ a self-reciprocal polynomial of degree $2d_i$, the polynomial $h_j$ is of degree $e_j$ and $*$ denotes
reciprocation. If $n$ is odd, then let $g_1=x+1$. The number of LCD index-3 negacirculant codes over $\F_q$ of length $3n$ and 1-generator
$<[1,a,b]>$ is then

$$(q^2-q+\eta(-1))\prod_{i=2}^s [q^{2d_i}-(q^{d_i}+1)(q^{2d_i}-q^{d_i})] \prod_{j=1}^t(q^{4e_j}-q^{3e_j}+q^{e_j}) $$
if $n$ is odd, and
$$\prod_{i=1}^s [q^{2d_i}-(q^{d_i}+1)(q^{2d_i}-q^{d_i})] \prod_{j=1}^t (q^{4e_j}-q^{3e_j}+q^{e_j}) $$
if $n$ is even.

\end{prop}

\pf We use the Chinese Remainder Theorem (CRT) decomposition of $R(n,\F_q),$ again. Since we are counting LCD quasi-twisted codes of index 3 and
generator matrix of the shape $<[1,a,b]>$, we are reduced to counting codes of length 3 and dimension 1 over some extension $\F_Q$ of $\F_q$
with certain properties.

In the case $Q=q$ we are reduced to counting LCD codes of parameters $[3,1]$ over $\F_q.$ We can invoke Corollary \ref{App7} to obtain the factor
$q^2-q+\eta(-1)$ in the stated formula.

A factor $g_i(x)$ of degree $2d_i$ leads to counting \emph{self-orthogonal hermitian codes} of length $3$ over $\F_Q,$ where $Q=q^{2d_i}.$
Writing the generator matrix of such a code in the form $\langle[1,a,b]\rangle,$ we must count the solutions of the equation
$1+aa^{q^{d_i}}+bb^{q^{d_i}}=0$ (or equivalently, $a^{1+q^{d_i}}+b^{1+q^{d_i}}=-1$). Then by Corollary \ref{App2} the number $N$ of the
solutions of that equation is $(\sqrt{Q}+1)(Q-\sqrt{Q})$ where $Q=q^{2d_i}$ i.e. $N=(q^{d_i}+1)(q^{2d_i}-q^{d_i}).$ By complementation the
number of LCD codes of index 3 is then $Q^2-N.$

In case of \emph{ reciprocal pairs} $(h'(x),h''(x))$, there are two dual constituent codes of length 3 over $\F_Q$ that is,
$\langle[1,a',b']\rangle$ and $ \langle[1,a'',b'']\rangle^\perp $.
By the condition for self-orthogonality we have to enumerate the cases when $[1,a',b']{\perp_E}[1,a'',b'']$ which
means counting the solutions of the equation $1+a'a''+b'b''=0$. Then, by Corollary \ref{App5} the number of the solutions of that
equation is $(Q^3-Q)$ where $Q=q^{e_j}.$ Thus, by complementation, the number of corresponding LCD codes is $Q^4-Q^3+Q.$
 \qed

The following will be used in our asymptotic study and it is a consequence of Proposition \ref{enum2}.

\begin{cor}\label{enumb}
 Let $q$ be odd, and $n$ be a power of $2$. If  $x^n+1$ factors as a product of two irreducible polynomials over $\F_q$, then the number of LCD three-negacirculant codes over $\F_q$ of length $3n$ is $q^{2n}-q^{\frac{3n}{2}}+q^{\frac{n}{2}}$.
\end{cor}

\subsection{Index $t>3$}

For higher indices, we do not have exact enumeration formula as in Propositions \ref{enum1} and \ref{enum2}. However, we have the following
analogue of Corollaries \ref{enuma} and \ref{enumb}, which is enough for asymptotic purposes in the next Section.

\begin{prop}\label{enumc}
 Let $q$ be odd, and $n$ be a power of $2$. If  $x^n+1$ factors as a product of two irreducible polynomials over $\F_q$, then the number of
 LCD $t$-negacirculant codes over $\F_q$ of length $tn$ is $q^{n(t-1)}-\left( q^{\frac{n}{2}(2t-3)}-\eta((-1)^{t-1})q^{\frac{n}{2}(t-2)}\right)$.
\end{prop}

\pf We use the Chinese Remainder Theorem (CRT) decomposition of $R(n,\F_q),$ as explained in $\S4$. There are two LCD codes of length $t$ over $\F_q$, $C'=\langle[1,a'_1,a'_2,...,a'_{t-1}]\rangle $ and $C''=\langle[1,a''_1,a''_2,...,a''_{t-1}]^{\perp}\rangle$. For LCD condition we need to have both $C' \cap C''^\bot$ and $C'' \cap C'^\bot$ are trivial. This boils down to the condition:
 $$a'_1a''_1+a'_2a''_2+...+a'_{t-1}a''_{t-1}\neq -1.$$
Now we need to count the solution of the above equation then by Corollary \ref{App5} the number $N$ of the solutions of that equation is $Q^{2t-3}-\eta((-1)^{t-1})Q^{t-2}$
where $Q=q^{\frac{n}{2}}.$ Thus by complementation, the number of corresponding LCD codes is $Q^{(2t-2)}-(Q^{2t-3}-\eta((-1)^{t-1})Q^{t-2})$. Hence,
$$N=q^{n(t-1)}-\left( q^{\frac{n}{2}(2t-3)}-\eta((-1)^{t-1})q^{\frac{n}{2}(t-2)}\right).$$ \qed

\section{Asymptotics}

In this section, we assume that $x^n+1,$ for $n$ a power of $2,$ has only two irreducible factors, say $h'(x)$ and $h''(x)$, and that they are
reciprocal of each other. Thus, $x^n+1=h'(x)h''(x).$ For convenience, let $K'=\frac{\F_q[x]}{\langle h'(x)\rangle}$ and
$K''=\frac{\F_q[x]}{\langle h''(x)\rangle}.$ By part (b) of both Theorems \ref{1} and \ref{2} (with $A=2$ and $A=1$, respectively), we obtain
such reciprocal pair of irreducible polynomials.

\begin{lem}\label{cover2}
 If $u \neq 0$ has Hamming weight $< n$, there are at most $(q^{\frac{(t-1)n}{2}}-1)$ polynomials with $x$-expansion $a_1,a_2,...,a_{t-1}$
 such that $u \in \C_{a_1,a_2,...,a_{t-1}}=<[1,a_1,a_2,...,a_{t-1}]>,$
and $\C_{a_1,a_2,...,a_{t-1}}$ is LCD.
\end{lem}

\pf Let $\C_{a_1,a_2,...,a_{t-1}}=<[1,a_1,a_2,...,a_{t-1}]>$, and let $u=(b_1,b_2,...,b_{t-1},b_t),$ with $b_1,b_2,...,b_{t-1},b_t$ vectors of
length $n$. The condition $u \in \C_{a_1,a_2,...,a_{t-1}}$ is equivalent to the
equations,
$$b'_{i+1}=a'_{i}b'_1  \hskip8mm \textrm{over} \hskip3mm K,' \hskip12mm \textrm{for all} \hskip3mm i=1,...,t-1$$
$$b''_{i+1}=a''_{i}b''_1   \hskip8mm \textrm{over} \hskip3mm K'', \hskip10mm \textrm{for all} \hskip3mm i=1,...,t-1$$
where $(b'_1,b''_1),$ denotes the image by the CRT in $K'\times K''$ of
the polynomial with $x$-expansion $b_1.$ Then we have, since $\C_{a_1,a_2,...,a_{t-1}}$ is LCD, that $$\langle[1,a'_1,a'_2,...,a'_{t-1}]\rangle \bigcap
\langle[1,a''_1,a''_2,...,a''_{t-1}]\rangle^\perp = \{0\},$$ which implies $a'_1a''_1+a'_2a''_2+...+a'_{t-1}a''_{t-1}\neq -1.$ \\

For all $i=1,2,...,t-1$:\\
\begin{itemize}
\item[(i)]If $b'_1 \neq 0$, then $a'_{i} = \frac{b'_{i+1}}{b'_1}$ has a unique solution.\\
\item[(ii)]If $b'_1 = 0$, then

(a)If $b'_{i+1} \neq 0$, then we have no solution.

(b)If $b'_{i+1} = 0$, then $a'_i$ is undetermined i.e. we have
$q^{\frac{(t-1)n}{2}}-1$ choices for $a'_i$ for all $i$.
\end{itemize}

Similarly, we have the same solutions for $a''_i$ for all $i$.
Therefore, for given $u=(b_1,b_2,...,b_{t-1},b_t),$ there are at most $q^{\frac{(t-1)n}{2}}-1$
choices for $a_i$ for all $i$. \qed

Recall the $q-$ary {\em entropy function} defined for $0<y< \frac{q-1}{q}$ by $$ H_q(y)=y\log_q(q-1)-y\log_q(y)-(1-y)\log_q(1-y).$$

\begin{thm} \label{re2}
 If $q$ is odd integer, and $n$ is a power of $2$, then, for any fixed integer $t\ge 2,$ there are infinite families of LCD index $t$
 negacirculant codes  of relative distance $\delta$ satisfying  $H_q(\delta)\geq \frac{t-1}{2t}$.\\

\end{thm}

\pf The negacirculant codes of index $t$ containing a vector of weight $d\sim t\delta n$ or less are by standard entropic estimates and Lemma
\ref{cover2} of the order $(q^{\frac{(t-1)n}{2}}-1) \times q^{tn H_q(\delta)}$, up to subexponential terms. This number will be less than the
total number of negacirculant codes of index $t$ which is, by Proposition \ref{enumc}, of the order of $(q^{2(t-1)\frac{n}{2}})= q^{(t-1)n}$.
\qed

\section{Conclusion}
In this paper, we have studied LCD quasi-twisted codes of index $t,$ where $t \geq 2$ emphasizing the aspects of enumeration for $t=2$ and
$t=3,$ and, for fixed $t,$ asymptotic performance. It is an open problem to derive exact enumeration formulas for $t>3.$ It is also an open
question to study the  asymptotic performance of quasi-twisted codes with more than one generator in their module structure.

\noindent{\bf \large{Acknowledgments:}} Patrick Sol\'{e} thanks Prof. Wolfmann for helpful discussions. G\"{u}neri and \"{O}zkaya are supported
by T\"{U}B\.{I}TAK project 215E200, which is associated with the SECODE project in the scope of CHIST-ERA Program. Sol\'{e} is also supported by
the SECODE project.


\section{Appendix}

\subsection{Norm function}
For all $x \in \F_{q^n},$ the norm of $x$ over $\F_q$ is a map
$Norm:\F_{q^n} \rightarrow \F_q$ defined by
$$Norm(x)=x^{(q^n-1)/(q-1)}.$$
Moreover, Norm is a multiplicative homomorphism which is surjective
(\cite[Theorem 2.28]{LN}). $Norm(0)=0$, so it maps $\F_{q^n}^*$ onto
$\F_{q}^*$, where each nonzero element in $\F_{q}^*$ has a preimage
of size $(q^n-1)/(q-1)$ in $\F_{q^n}^*$.

Hence, for $n=2$, we have $Norm(x)=x^{1+q}$ for all $x \in
\F_{q^2}^*.$ It is a $(q+1)$ to 1 map.

\begin{cor} \label{App2}
If $q$ is odd, and $n$ is coprime with $q$, then the number of
solutions $(a,b)$ in $\F_{q^2}$ of the equation
$a^{(1+q)}+b^{(1+q)}=-1$ is $(q+1)(q^2-q)$.
\end{cor}

\pf We could invoke \cite[Corollary 4]{W}, with $b^n = 1,\, \eta = (?1)^{ t/r+1} = 1$ , but we prefer to give a self-contained argument. If
$b^{1+q}=-1$, then $a=0$. By the norm map, there are $(1+q)$ such $b \in \F_{q^2}^*.$ Then, we have $q+1$ solutions in  $\F_{q^2}^*$ of this
form. If $b^{1+q} \neq -1$, then $a^{1+q}= -1-b^{1+q}$ has $(q+1)$ distinct solutions by the norm and $b^{1+q} \neq -1$ is true for
$(q^2-(q+1))$ elements in $\F_{q^2}$. Therefore, we have $(q^2-q-1)(q+1)$ solutions for this case. Hence, in total we have
$(q+1)+(q^2-q-1)(q+1)= (q+1)(q^2-q)$ solutions in $\F_{q^2}^*.$ \qed

\subsection{Quadratic Forms}

We quote Theorem 6.26 of \cite{LN}.
Define the function $v$ as $v(x)=-1$ if $x$ is nonzero and $v(0)=q-1.$
{\thm Let $f$ denote a quadratic form in an even number $n$ of variables over $\F_q,$ with $q$ odd. Denote by $\Delta$ the discriminant of $f.$ Given $b\in \F_q,$ the number of solutions in $(x_1,\dots,x_n) \in \F_q^n $ of
$$f(x_1,\dots,x_n)=b   $$ is $$q^{n-1}+v(b)\eta((-1)^{\frac{n}{2}}\Delta)q^{\frac{n-2}{2}}. $$
}
 From this general statement, we derive two results  useful for our purposes.

\begin{cor} \label{App7}
If $q$ is odd, then  the number of
solutions $(x,y)$ in $\F_{q}$ of the equation $x^{2}+y^{2}=-1$ is
$$ q-\eta(-1).  $$
\end{cor}

\pf Follows by the previous Theorem with $b=-1,\,n=2,\,f=x^2+y^2,\, \Delta=1.$
\qed

Recall that $\eta(-1)=1$ if and only if $q$ is a square or if $q$ is not a square, but the characteristic of $\F_q$ is $\equiv 1 \pmod{4}.$
\begin{cor} \label{App5}
The number of solutions of $x_1y_1+x_2y_2+...+x_{t-1}y_{t-1}=-1$ is $$q^{2t-3}-\eta((-1)^{t-1})q^{t-2}.$$
\end{cor}

\pf
Letting $A_1=x_1+y_1$, $A'_1=x_1-y_1,$ $A_2=x_2+y_2$, $A'_2=x_2-y_2,...,$ $A_{t-1}=x_{t-1}+y_{t-1}$, $A'_{t-1}=x_{t-1}-y_{t-1},$  the above equation can be cast into the following diagonal form.\\
$${A_1}^2-{A'_1}^2+{A_2}^2-{A'_2}^2+...+{A_{t-1}}^2-{A'_{t-1}}^2=-4$$
Now we can apply the above Theorem with $n=2(t-1),\,b=-4,\,\Delta=1,$ to obtain the stated result.
\qed

\end{document}